\documentclass{l4dc2026}

\title[Coordination of Heterogeneous Teams with Uncertain Task Requirements]{Coordination of Heterogeneous Teams\\with Uncertain Task Requirements}
\usepackage{times}
\usepackage{booktabs}
\usepackage{enumitem}
\usepackage{wrapfig}
\usepackage{url}

\coltauthor{\Name{Jongann Lee} \Email{jongann2@illinois.edu} \\
 \Name{Melkior Ornik} \Email{mornik@illinois.edu}\\
 \addr University of Illinois Urbana-Champaign,  Urbana, IL, USA, 61801}

\begin{document}

\maketitle

\begin{abstract}%

Coordinating a team of heterogeneous agents to complete tasks with unknown requirements is a challenging multi-agent planning problem. Agents differ in their ability to discover task requirements and in their ability to engage tasks, so the planner must coordinate task discovery, execution, and staging. This paper introduces a hierarchical planner that uses a graph neural network policy to assign actions to agents, then finds a path to the action's location using a shortest-path algorithm. The central part of our approach is the CAAT graph, which is a graph representation of the planning context consisting of the agents, tasks, capabilities, and actions. We demonstrate that our learned planner can jointly coordinate a heterogeneous team to simultaneously discover task requirements and complete tasks in a shorter time compared to baselines that optimize task discovery and task engagement independently. Furthermore, we show that our policy can zero-shot plan on out-of-distribution settings by training only on an outdoor map then transferring without finetuning to an indoor map with more agents, tasks, and capabilities than seen during training. Code is available at { \smaller \url{https://github.com/jongann-lee/Heterogeneous_Capability_Planning/tree/submission1}}.

\end{abstract}


\section{Introduction} \label{sec:Introduction}

Envision a scenario where satellite imagery has identified threats at known locations across mountainous terrain. A heterogeneous team is tasked with eliminating these threats. Although the target locations are known, the capabilities required to eliminate each target remain uncertain. Some members possess specialized engagement capabilities but have limited vision, like a bomb defuser wearing a bulky mask. Others are lightly equipped scouts; they can identify threats from distant viewpoints by traveling to vantage points (like the top of a hill), but cannot eliminate them. Committing an agent to an incompatible task can result in its loss, hampering the team's ability to complete the objective. 

Coordinating both task requirement discovery and task engagement presents numerous challenges. Attackers have to choose between moving to a known task far away or a closer unknown task. Scouts have to consider the utility of revealing a task with respect to the current attacker states. Hybrid scout-attackers can either reveal unknown task requirements or complete a known compatible task. Thus, coordinating such a heterogeneous team requires more than optimizing the scouts and workers independently.

Previous research has largely focused on coordinating a homogeneous team with identical task requirements ~\citep{Choi2009CBBA, liu2011MRTA_hungarian}. More recently, multi-agent planning methods that consider different agent capabilities and unknown task requirements have been proposed ~\citep{Long2026IAOM, meighan2026HALO, Rossano2026AURA}. However, no previous work has built a multi-agent planner that can accommodate a varying set of capabilities (including the ability to scout) and uncertain task requirements that are revealed by sending a scout to a location that can see that task. 

Our approach is to create a graph representation of the planning context, which is then used by a graph neural network (GNN) policy to allocate agents to actions. We introduce \textbf{CAAT graph}, a graph comprising \textit{agent, task, capability,} and \textit{action} nodes that captures the key decision-making variables and their relations. The action nodes represent four types of actions: \textit{execute task, observe, stage,} and \textit{wait} and gives the planner a consistent set of actions on different maps and at different time steps. Using the CAAT graph as the representation allows a pretrained policy to transfer to a different map with new agent, task, and capability settings without any retraining. Our planner is hierarchical; the GNN planner specifies the \textit{action} and a low level path planner finds the shortest path to that action's location. We pretrain the GNN policy on a large set of simulated rollouts on a single map using reinforcement learning (RL). Our contributions are as follows:
\begin{enumerate}[itemsep=2pt, parsep=2pt]
    \item We define the heterogeneous team multi-robot task allocation (MRTA) problem with uncertain task requirements. Our problem is an extension of MRTA in two parts: (i) a \textit{capability} that each agent can possess and that each task requires but is unknown a priori; (ii) a \textit{visibility mapping} that defines the set of nodes from which the task requirement can be observed.
    \item We create a graph representation of the planning context called the CAAT graph, which provides a consistent set of inputs and actions for the planner. Experimental results demonstrate that using the CAAT graph enables a pretrained GNN policy to zero-shot transfer to out-of-distribution configurations and maps.
    \item We show that our method achieves a lower problem completion time than a baseline that optimizes scouting and task execution independently, demonstrating that our policy successfully exploits the benefits of jointly coordinating task requirement discovery and task execution.
\end{enumerate}

\section{Related Work} \label{sec:Related_Work}

\paragraph{Multi-Robot Task Allocation (MRTA):} The goal of MRTA is to coordinate a team of agents to complete a set of tasks while minimizing the cost ~\citep{Gerkey2004MRTA}. Many solutions to MRTA have been proposed, including optimization-based approaches ~\citep{koes2005heterogeneous, liu2011MRTA_hungarian, shorinwa2023ADMM_MRTA}, auction-based approaches  ~\citep{gerkey2002MRTA_auction, Choi2009CBBA}, and heuristics ~\citep{parker1998alliance}. More recently, deep learning has been applied to MRTA using attention mechanisms, GNNs, and RL ~\citep{paul2022CapAM, wang2022ScheduleNet, altundas2025TargetNet}. Typically, a GNN processes a graph of agents and tasks to create an embedding, which is then used by a decoder to create a probability distribution of tasks for each agent. The policy is pretrained on a large set of sampled rollouts using RL. While these approaches are not directly applicable to our problem, our method builds on the previously established GNN + decoder + RL method.

\paragraph{Watchman Routing Problem (WRP):} The scout's objective in our problem is to observe all of the tasks on the map. WRP is a closely related problem, where the objective is to find the shortest path on which every node is observed at least once ~\citep{chin1986WRP_def}. WRP on grid maps has been solved by formulating it as a search problem on a tree then using $A^*$ with a heuristic ~\citep{seiref2020WRP_single, livne2023WRP_multi}. These methods are too narrowly focused on just scouting; we use this method as part of our baseline and demonstrate that our method is better than it. 

\paragraph{Heterogeneous Team MRTA with Uncertain Task Requirements:} Recent works have studied coordinating a heterogeneous team for MRTA with uncertain task requirements. IAOM uses a GNN with RL to coordinate a homogeneous team of robots to solve a variant of MRTA with an unknown number of robots required per task ~\citep{Long2026IAOM}. AURA is an auction based method that coordinates a team of search robots and debris robots to complete a heterogeneous set of search and debris tasks ~\citep{Rossano2026AURA}. \citet{Beck2016online_SAR} alternates search-route and rescue-schedule optimization using sampled task realizations for search and rescue. None of the previous works consider hybrid agents capable of both scouting and engaging tasks, variations in visibility, and a variable number of capabilities.

\section{Problem Formulation} \label{sec:Problem_Formulation}

Our problem is motivated by the problem of coordinating a team of agents with different capabilities to complete a set of tasks with unknown requirements that are revealed by the scout agents. We formulate the problem of \textit{heterogeneous team MRTA with uncertain task requirements}. A directed, weighted graph $G = (\mathcal V, \mathcal E)$ represents a map, with node $v \in \mathcal V$ for location, and $e \in \mathcal E $ indicating traversability between two nodes. Let $c(v_1, v_2), (v_1, v_2) \in \mathcal E$ be the traversal time of an edge. We define the visibility map $\mathrm{vis}(v) \subset \mathcal V$  as the set of nodes visible from node $v$ \citep{jongann_iros26}. Both $c(v_1, v_2)$ and $\mathrm{vis}(v)$ are known a priori. The traversal time is directional; e.g. traveling uphill from node $v_1$ to $v_2$ takes longer than traveling downhill from $v_2$ to $v_1$.

There are multiple agents $A_i, 1 \leq i \leq N_a$ and multiple tasks $T_j, 1 \leq j \leq N_t$. Tasks are stationary and their locations are known. Agents traverse from node $v_1$ to node $v_2$ along an edge, with their traversal time corresponding to the traversal cost of the edge $c(v_1, v_2)$. Two or more agents can occupy the same node simultaneously. 

We define a set of capabilities $C_k \in \mathcal C, 0 \leq k \leq N_c$. Intuitively, a capability corresponds to the \textit{trait} that an agent needs to have to complete a task. The \textit{requirement} of a task $C(T_j) \in \mathcal C \backslash \{C_0 \} $ is the capability required to complete task $T_j$. The task requirements are initially unknown with the prior probability of $T_j$ requiring $C_k$ being $p(C(T_j) = C_k)$. An agent can possess multiple capabilities $C(A_i) \subset \mathcal C$. Our problem is an ST-SR-TA MRTA problem ~\citep{Gerkey2004MRTA}. $C_0$ represents the \textit{scout} capability. It is only given to agents, and only agents with $C_0$ can observe an unknown task's required capability within their current visibility $\mathrm{vis}(v)$. Upon reaching a node, a scout instantly reveals the exact capability requirements of all tasks visible from that node at no additional observation cost. We assume that this information is instantly available to all agents.

An agent completes a task if it reaches it and possesses the capability required by the task. An agent can only be assigned to complete a task if its requirements are known and possessed by the agent. The objective is to minimize the makespan: the time to complete all the tasks.

\section{Method} \label{sec:Method}

\begin{figure}[t]
    \centering
    \includegraphics[width=\linewidth]{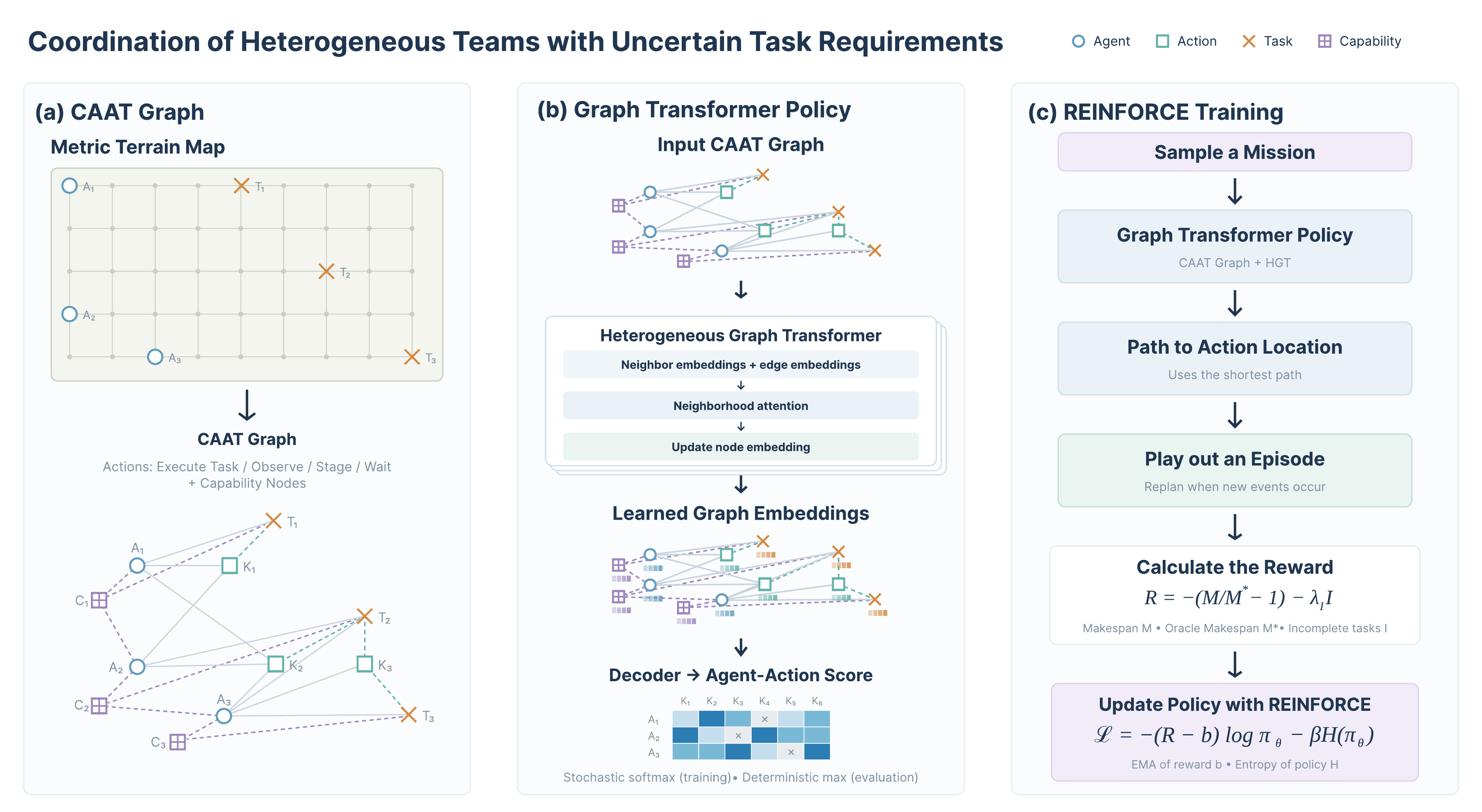}
    \vspace{-25pt} 
    \caption{We solve the heterogeneous team MRTA with uncertain task requirements by creating a graph representation of the planning context called CAAT graph. A GNN then learns graph embeddings on the CAAT graph and a decoder outputs a probability distribution of actions for each agent. The policy is pretrained with RL using simulated rollouts.}
    \vspace{-15pt} 
    \label{fig:main_method_diagram}
\end{figure}

The problem as defined in Section \ref{sec:Problem_Formulation} consists mainly of two parts: task requirement discovery and task execution. Each part by itself has an optimal solution: discovering all of the task requirements is equivalent to WRP ~\citep{livne2023WRP_multi} and task execution once the task requirements are known is a well defined optimization problem ~\citep{koes2005heterogeneous}. The difficulty is in combining both: task requirements should be discovered in a way that reduces task execution time, non-scout agents should move to staging locations to quickly engage newly scouted tasks, and hybrid agents have to determine when to scout and when to engage a known task. While the problem can be formulated as a centralized partially observable Markov decision process (POMDP), getting an optimal POMDP solution is intractable due to the large state space.

Our solution to the problem defined in Section \ref{sec:Problem_Formulation} is a hierarchical planner as outlined in Figure \ref{fig:main_method_diagram}. A high-level GNN planner decides which \textit{action} each agent should take. As the input for this planner, we create a graph representation of the planning context called the \textbf{CAAT graph}. Once an action has been assigned, the agent's path to the action's location is determined using a shortest path algorithm such as $A^*$. The agents follow their planned paths until an agent completes its intended action or an unknown task requirement is revealed, at which point the planner replans. This hierarchy reduces the decision space that must be learned and preserves the shortest path optimality for routing. The GNN planner is pretrained on simulated rollouts using RL.

\subsection{CAAT Graph} \label{subsec:CAAT_graph}


\begin{figure}[t]
    \centering
    \includegraphics[width=\linewidth]{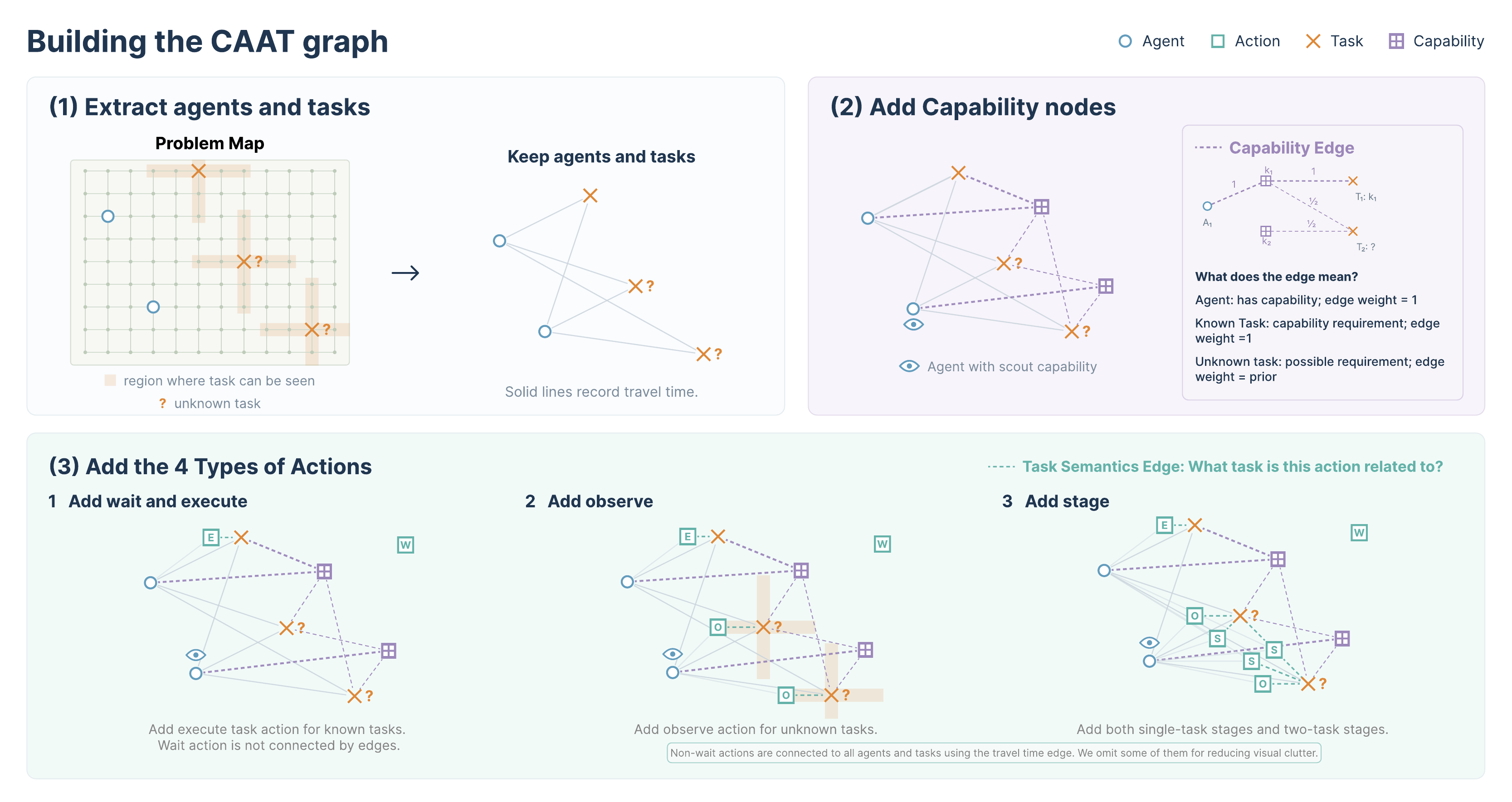}
    \vspace{-15pt} 
    \caption{An illustration of how the CAAT graph is made. See Subsection \ref{subsec:CAAT_graph} for more details.}
    \vspace{-15pt} 
    \label{fig:CAAT_Graph_step}
\end{figure}

The Capability-Agent-Action-Task graph, or CAAT graph, represents the planning context at a given time. It consists of four types of nodes and three types of edges as seen in Figure \ref{fig:main_method_diagram}. Its goal is to provide a consistent and compressed representation of the situation and actions. We explain the CAAT graph by following its creation step by step as in Figure \ref{fig:CAAT_Graph_step}.

First, we extract the agents and tasks from a given timestep of the problem. Every task is connected to every agent with a travel time edge. A connection via a travel time edge represents a connection that may be traversed by an agent. The edge feature is the shortest path travel time, which is then normalized using z-score normalization. The task features are the position (normalized to $[0,1]^2$) and two Boolean variables: task completion and whether the required capability is known. The agent features are the Boolean indicating whether the agent is alive, remaining transit time, a Boolean indicating whether the agent has the scout capability $C_0$, and the agent's position (normalized).

We then add capabilities as graph nodes. Each capability is a node, and it is connected to agents and tasks via a capability edge. If an agent has a capability, then it is connected via a capability edge (with edge weight 1) to the corresponding capability node. Tasks with known requirements are also connected to their corresponding capability node with a capability edge of weight 1. For a task $T_j$ with unknown requirements, the capability edge connects all of the possible requirements with the prior $p(C(T_j) = C_k)$ as the edge weight. Representing capabilities as nodes allows a pretrained GNN policy to accept additional capabilities without changing the model dimensions. The scout capability is an exception; it is encoded as an agent feature.

Finally, we add four types of actions to the graph. 
\begin{enumerate}[itemsep=2pt, parsep=2pt]
    \item \textbf{Wait:} remain at the current location.
    \item \textbf{Execute:} go to a known task.
    \item \textbf{Observe:} reach the closest suitable location to observe a subset of unknown tasks. 
    \item \textbf{Stage:} position near unknown tasks in anticipation of their requirements being revealed. 
\end{enumerate}
Each action is connected to its relevant tasks via a task semantics edge. Each action is also connected to all agents and all tasks via a travel time edge. The wait action is the exception: it is not connected to any other node since it is not related to any task and does not have a location on the map. More details are in Appendix \ref{app:CAAT_details}.

\subsection{Heterogeneous Graph Transformer Policy} \label{subsec:HGT_policy}

At each planning decision, we use a GNN to create embeddings of the CAAT graph. A brief introduction to GNNs and the Graph Attention Network (GAT) is in Appendix \ref{app:GNN_Intro}. We choose the heterogeneous graph transformer (HGT) as the GNN architecture. The CAAT graph has four different types of nodes and three different types of edges, so our HGT uses a different linear transformation for each direction of each edge type, which we denote as a relation type $r=$ (source node type, edge label, destination node type). Details on the HGT can be found in Appendix \ref{app:HGT_details}.

The decoder computes a score $\eta$ for each agent--action pair, then assigns actions to agents using the score. Using the final agent embedding $\mathbf x_{out}(A_i)$, final action embedding $\mathbf x_{out}(K_n)$ and the corresponding travel-time edge embedding $\mathbf z_{in}$, the score is
\begin{equation} \label{eq:agent_action_similarity}
    \eta(A_i,K_n)
    = f_\theta\left(
        \mathbf x_{out}(A_i) \Vert
        \mathbf x_{out}(K_n) \Vert
        \mathbf z_{in}
    \right),
\end{equation}
where $f_\theta$ is an MLP scorer and $\Vert$ denotes concatenation. Incompatible pairs such as (non-scout agent, observe) and (known task, agent without requirement) are assigned $\eta(A_i,K_n)=-\infty$.

The actions are sequentially assigned to agents. At assignment step $s$, let $\mathcal F_s$ contain the feasible and unassigned (agent, action) pairs. The score defines the probabilities of selecting each pair using softmax:
\begin{equation} \label{eq:policy_softmax}
    p_s(A_i,K_n)
    = \frac{\exp\left(\eta(A_i,K_n)\right)}
    {\displaystyle\sum_{(A_j,K_m)\in\mathcal F_s}
        \exp\left(\eta(A_j,K_m)\right)},
    \quad (A_i,K_n)\in\mathcal F_s.
\end{equation}
Actions that share the same location are merged to form one destination and the probability of selecting it is the sum of the probabilities of the component actions. During training, the decoder samples an agent and destination using these probabilities; during evaluation, it selects the feasible agent--destination pair with the highest probability. After each selection, the assigned agent and action are removed $\mathcal F_{s+1} = \mathcal F_s \backslash \{ (A_i, K_n): A_i = A^* \text{ or } K_n = K^* \}.$ Selection continues until no feasible pair remains. Additional decoder details are provided in Appendix~\ref{app:HGT_details}.

\subsection{Policy Pretraining} \label{subsec:policy_pretraining}

We pretrain the neural policy $\pi_\theta$ using a large number of simulated rollouts. During the rollout, the policy $\pi_\theta$ assigns agents to actions, and the agents follow the shortest path to the action's location. They continue on the path until an unknown task requirement is revealed, a task is completed, or an agent reaches its destination, at which point the planner replans. The rollout ends either when all tasks are complete, or when all agents select wait and the rollout stalls.

At the end of each rollout, we calculate the rollout reward by comparing it to the \textit{oracle}, an optimal planner that knows the ground-truth task requirements. The primary component in the reward is the \textit{regret} which is the makespan of our policy's rollout $M$ divided by the makespan of the oracle $M^*$. We add a penalty term for the number of incomplete tasks $I$ to discourage stalling, with $\lambda_I$ as the hyperparameter.
\begin{equation} \label{eq:episode_reward}
    R = - \left( \frac{M}{M^*} - 1\right) - \lambda_I I
\end{equation}

Like previous works ~\citep{kool2018attention}, we use the REINFORCE algorithm ~\citep{williams1992REINFORCE} to update the weights.  To further stabilize training, we use an exponential moving average $b$ of the reward as a baseline ~\citep{sutton1998RL}. An entropy term $H(\pi_\theta)$ is also added to encourage the policy to explore ~\citep{haarnoja2018sac}. The weights are updated using the following loss:
\begin{equation} \label{eq:REINFORCE_loss}
    \mathcal L = - (R-b) \log \pi_\theta - \beta H(\pi_\theta).
\end{equation}
Additional details on policy pretraining and the oracle are in Appendix \ref{app:policy_pretraining} and \ref{app:baseline_planner}.

\section{Experiments} \label{sec:experiments}

\subsection{Experiment Setup} \label{subsec:experiment_setup}

\paragraph{Maps.} We use three maps in our experiments: WV, WA, and KTH. WV and WA are outdoor maps of locations in the US states of West Virginia and Washington, and are created using the topography data from OpenTopography ~\citep{OpenTopography}. The maps are discretized into a $64 \times 64$ grid graph. The traversal cost is calculated using Tobler's hiking function with road edges having $40\%$ lower traversal cost ~\citep{tobler1993three}. The visibility is modeled using viewshed analysis ~\citep{Toma_Zhuang_Richard_Metz_2026}. The KTH map is an indoor map of a building at KTH Royal Institute of Technology ~\citep{ho2025mapex}. It is also discretized into a grid; its traversal cost is the edge length and the visibility is modeled using line of sight. WV is used as the training map, WA is used as the in-distribution test map, and KTH is used as the out-of-distribution test map with 4.6 times as many nodes compared to WV or WA.

\paragraph{Capabilities, Agents, and Tasks.} We pretrain our policy on the WV map with 3--6 agents and 5--9 randomly placed tasks. There are 3 non-scout capabilities, and they are assigned randomly such that all capabilities are matched with at least one agent. All agents start at the same node. For testing, we use a fixed suite of capabilities, agents, and tasks for each map. All maps have three sets of agent capabilities and three sets of task requirements for each number of agents $N_a$ and tasks $N_t$. For WV and WA, $3 \leq N_a \leq 6$ and $5 \leq N_t \leq 9$ so there are a total of 12 agent configurations and 15 task configurations. We test every pair of agent and task configurations, resulting in a total of 180 test problems per map for WV and WA. The KTH map is tested with 8 agents and 14 tasks for a total of 9 problems. 

\paragraph{Baselines.} We compare the performance of CAAT Graph + HGT with three baselines. The `oracle' is an optimal planner that knows all of the task requirements, and its makespan is the lowest achievable makespan for a given problem. `Scout then engage' first discovers all of the task requirements using an optimal WRP planner, then uses the oracle to execute all of the tasks. `Engage while scout' uses the same WRP planner to coordinate scouts and simultaneously assigns non-scouts to execute known tasks using the oracle.  Additional details on the baselines are in Appendix \ref{app:baseline_planner}.

\paragraph{Pretraining.} Our policy is pretrained on 12,800 sampled rollouts on the WV map. On a desktop with an NVIDIA RTX 5090 GPU, pretraining takes approximately 8 hours. Once pretrained, we freeze all policy weights and do not finetune them during evaluation. Additional experimental settings are provided in Appendix~\ref{app:detail_on_results}.

\subsection{Results} \label{subsec:results}

This subsection compares the performance of each policy on each map. We choose to compare regret $\frac{M}{M^*}$ rather than makespan because a problem with more agents and fewer tasks requires less time to complete regardless of policy. We first evaluate on the training map WV, to evaluate whether our policy learns to plan effectively within the training domain. We then evaluate on unseen maps WA and KTH. WA introduces a different map with a different starting position, but follows the same terrain traversal rule, viewshed visibility, and the same ranges of capability, agent and task counts. The KTH map has the largest deviations, with a different traversal model, different visibility, and a larger number of capabilities, agents, and tasks. 

\begin{figure}[t]
    \centering
    \includegraphics[width=0.9\linewidth]{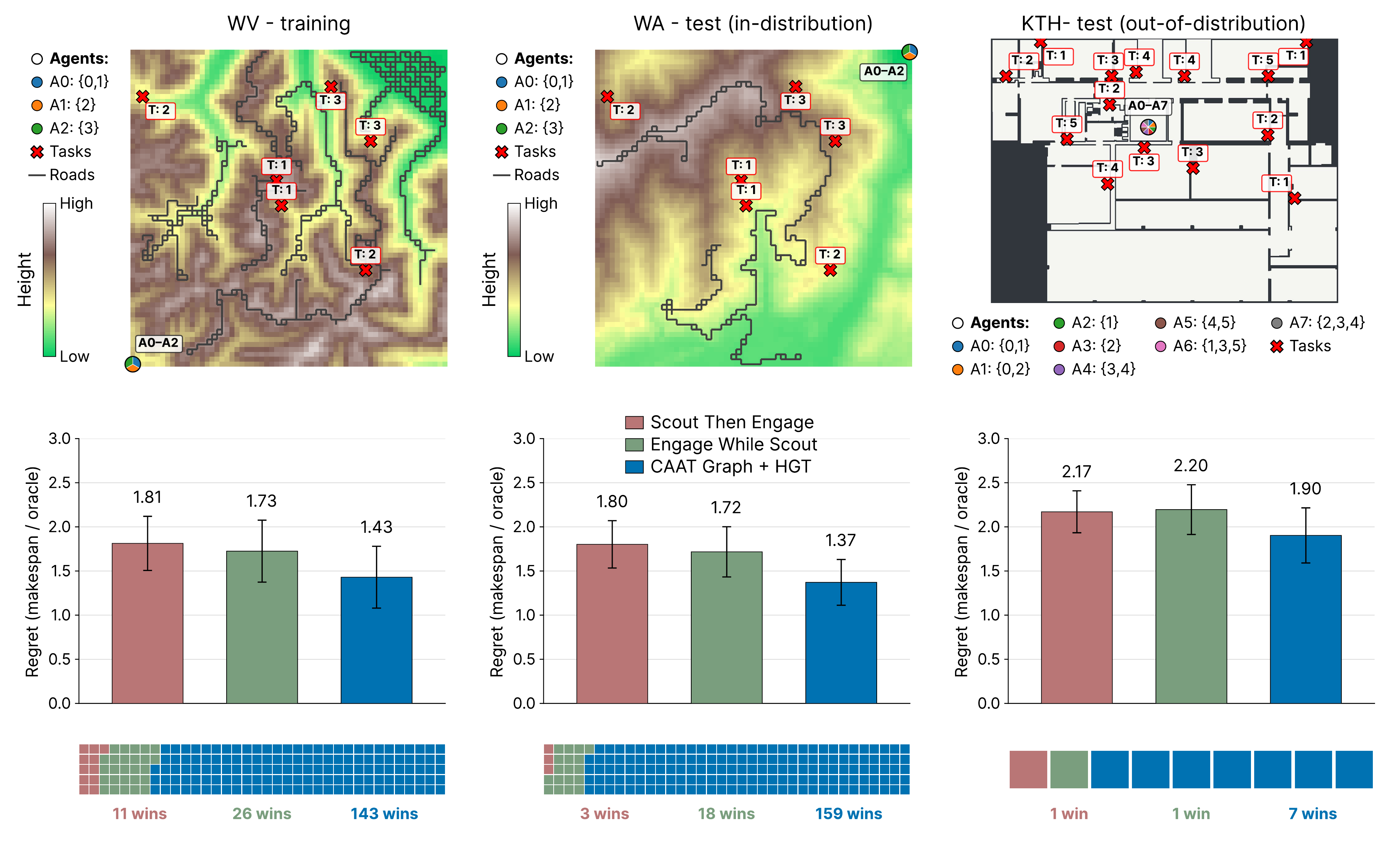}
    \vspace{-5pt} 
    \caption{Each column represents results from one map. WV is the training map, WA is the in-distribution test map, and KTH is the out-of-distribution test map. The bar graph in the middle is the average regret across all scenarios and the bottom diagram shows which policy (excluding oracle) had the lowest regret for each test problem.}
    \vspace{-15pt} 
    \label{fig:main_regrets}
\end{figure}

The average regret of each policy is shown in Figure \ref{fig:main_regrets}. On all three maps our pretrained CAAT Graph + HGT has a lower regret compared to the two baselines. On the WV and WA maps, our method results in $17.3\%$ and $20.3\%$ lower average regret respectively compared to the best baseline, demonstrating that our method can jointly coordinate task requirement discovery and task completion. Even on the KTH map, which is an unseen map with more agents, tasks and capabilities than seen during training, the pretrained CAAT Graph + HGT results in a $12.4\%$ lower regret compared to the best baseline. The results on the KTH map show that our method zero-shot transfers to out-of-distribution settings without any finetuning. Furthermore, our method is better than both baselines for more than $77\%$ of the test problems regardless of the test map, demonstrating that it is able to improve across a wide range of problems.

\begin{figure}[t]
    \centering
    \begin{minipage}[t]{0.59\linewidth}
        \vspace{0pt}
        \centering
        \includegraphics[width=\linewidth]{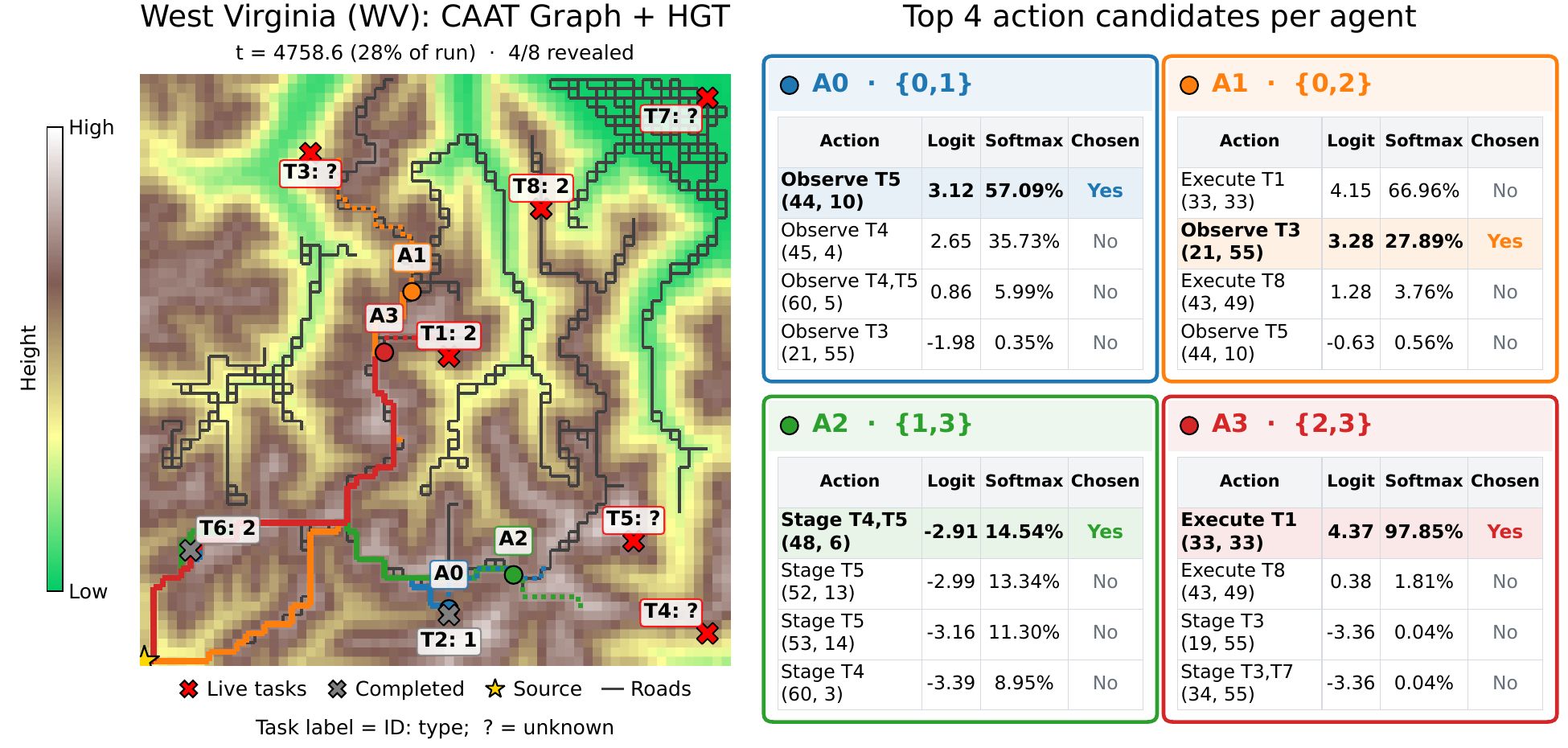}
        \vspace{-15pt} 
        \caption{A planning decision on WV (left) and the top 4 actions for each agent (right).}
        \vspace{-15pt} 
        \label{fig:detailed_decision}
    \end{minipage}\hfill
    \begin{minipage}[t]{0.39\linewidth}
        \vspace{0pt}
        \centering
        \includegraphics[width=\linewidth]{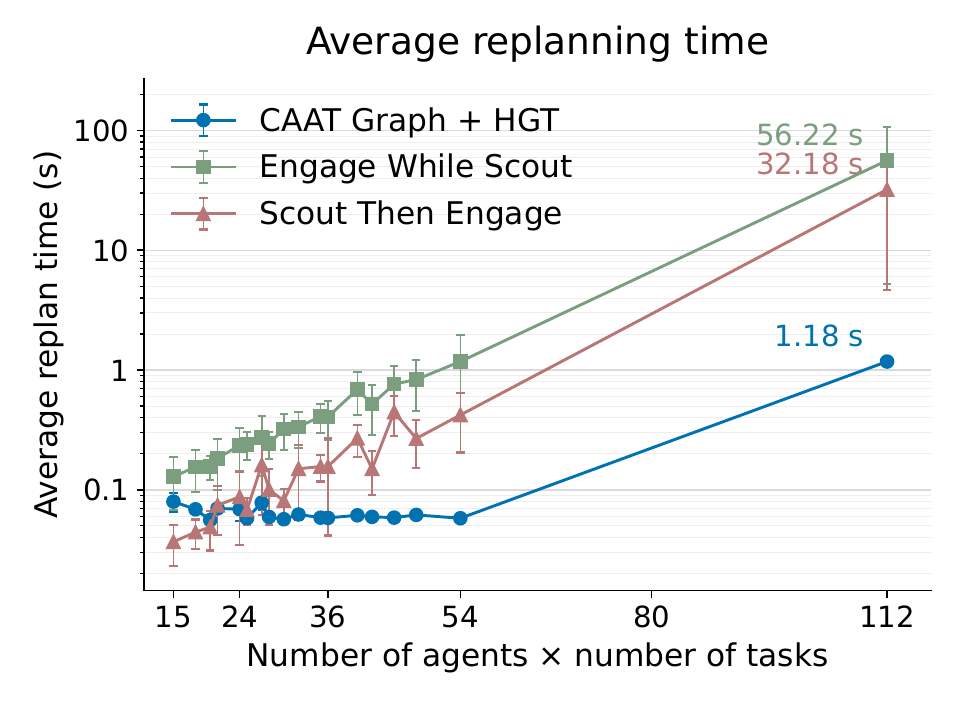}
        \vspace{-15pt} 
        \caption{Average time per planning decision.}
        \vspace{-15pt} 
        \label{fig:planning_time}
    \end{minipage}
\end{figure}

Next, we examine an actual planning decision produced by our method. Figure \ref{fig:detailed_decision} shows a planning decision on WV with 4 agents and 8 tasks (2 completed, 2 known but incomplete, 4 unknown). In the middle of the map, agents 1 and 3 are close to known task 1 and unknown task 3. Both agents have the capability required to complete task 1, so both of them get a large score for completing task 1. However, only agent 1 has the scout capability, so agent 3's score for completing task 1 is higher, and our method assigns executing task 1 to agent 3 and assigns agent 1 to observe task 3. This intuitively makes sense: if both agents can complete task 1, but only one can observe task 3, then the scout should go observe. At the bottom, agents 0 and 2 are close to two unknown tasks. Therefore, agent 0 with the scout capability gets large scores for observing task 4 or task 5 and agent 2 gets large scores for staging for those two tasks. These decisions show that our method makes intuitively correct decisions for the given context of the problem.

The average time per planning decision with respect to the product $N_a N_t$ is plotted in Figure \ref{fig:planning_time}. Note that most of the data comes from WV and WA with 4k nodes, but the data for $N_aN_t = 112$ comes from the larger KTH with 18k nodes. Engage while scout is consistently the slowest planner. Scout then engage is faster than our method at low $N_aN_t$ but gets significantly slower as the number of agents and tasks increases. In the KTH test, our method plans in just 1.18 seconds, compared to 32.18 and 56.22 seconds for the baselines. This speedup is achieved by replacing the baselines' combinatorial search over the task assignment and visitation sequences with a neural policy $\pi_\theta$ that assigns high level actions on the CAAT graph.

\subsection{Ablation} \label{subsec:ablation}

\begin{table}[htbp]
    \centering
    \small
    \setlength{\tabcolsep}{4pt}
    \begin{tabular}{@{}lc||cccccc@{}}
        \toprule
        Map & Default
            & \shortstack{Dot-product\\score}
            & \shortstack{Max\\normalization}
            & \shortstack{No\\position}
            & \shortstack{Single-task\\staging}
            & \shortstack{Single-task staging\\+ No position}
            & \shortstack{No\\staging} \\
        \midrule
        WV  & 1.43 & 1.65 & 1.42 & 1.51 & 1.63 & 1.46 & 1.74 \\
        WA  & 1.37 & 1.59 & 1.40 & 1.51 & 1.57 & 1.38 & 1.68 \\
        KTH & 1.90 & 2.99 & 1.96 & 2.11 & 2.10 & 2.08 & 2.19 \\
        \bottomrule
    \end{tabular}
    \vspace{-5pt} 
    \caption{Average regret $M/M^*$ for the default configuration and ablations.}
    \vspace{-5pt} 
    \label{tab:ablation}
\end{table}

We present a comprehensive ablation study by modifying various elements of our method. The results of the ablations are in Table \ref{tab:ablation}. We modify the following parameters. 

\paragraph{Dot product Score:} A simple way to score a pair of embeddings is to use their dot product. In this test we replace the MLP scorer with a dot product score $\eta(A_i, K_n) = \mathbf x_{out}(A_i) \cdot \mathbf x_{out}(K_n)$. The result is higher regret on all three maps, with especially higher regret on KTH. A possible factor is that the dot product decoder doesn't use $\mathbf z_{in}$ directly when outputting the score $\eta$. We therefore conclude that using the dot product scorer degrades our method and especially degrades its ability to transfer to unseen maps and settings.

\paragraph{Max Normalization:} Another common strategy for normalizing a set of real numbers is to divide by the maximum absolute value. For travel time, max normalization has the additional benefit of mapping non-negative travel times to $[0,1]$ while preserving zero. In this test the travel time on the CAAT graph is normalized by dividing by the largest travel time in the graph. Using max normalization yields marginally better results on WV but marginally worse results on WA and KTH compared to using z-score normalization. Therefore, we use z-score normalization instead of max normalization for the travel time edges.

\paragraph{No Position:} The high level decision of assigning actions to agents should only depend on the travel time between the various nodes. Removing the position could potentially make our method more robust by avoiding the possibility of overfitting to a single map. However, the results in Table \ref{tab:ablation} show that removing the position features increases the regret across all maps. This refutes the hypothesis that removing position will lead to a more robust planner.

\paragraph{Staging:} How much of the efficiency comes from the ability to stage? In this test `single-task staging' creates staging actions for single tasks; `no staging' means no staging actions are in the CAAT graph. Only having staging for a single task results in higher regret compared to having both single task and two task staging. Interestingly, the performance with only single task staging actually improves when the position is also removed as a feature. Removing the staging actions entirely significantly increases the average regret, as the planner is no longer able to send non-scouts to wait close to unknown tasks. Both our policy without staging and `engage while scout' simultaneously coordinate scouts and non-scouts using observation and execution actions. Our policy achieves comparable regret on WV and KTH and slightly lower regret on WA, demonstrating that it learns coordination competitive with a baseline even without staging.

\section{Limitations} \label{sec:limitations}

Our planner assumes that the tasks' locations are known in advance. In a multi-agent planning problem with uncertain task requirements the location of the task may also be uncertain. Additionally, our method is a centralized planner that assumes perfect communication and instant relay of newly acquired information. A decentralized method would be more robust to faulty communication and would scale better with a larger number of agents. 

\section{Conclusion} \label{sec:conclusion}

In this paper, we considered the heterogeneous team multi-robot task allocation problem with uncertain task requirements. To solve it, we created a graph representation of the planning context called CAAT (Capability-Agent-Action-Task) graph that captures the key decision-making variables and provides a consistent set of actions on different maps and settings. Our hierarchical planner assigns actions to agents by processing the CAAT graph using a graph neural network, and then generates the path to the action's location using a shortest-path algorithm. Our method outperformed two baselines which optimize task discovery and task execution independently, showing that it efficiently coordinates task discovery, execution and staging for unknown tasks. Furthermore, our policy trained on an outdoor map transferred without finetuning to a larger indoor map with more agents, tasks, and capabilities than seen during training while still outperforming the baselines.

\acks{
This work is supported by the Office of Naval Research grant no. N-00014-25-1-2369 and Office of Naval Research grant no. N-00014-25-1-2519.

Generative AI has been used to conduct literature search, to write code, to create figures, and to proofread the manuscript.
}

\bibliography{l4dc2026-sample}

\clearpage
\section{Appendix} \label{sec:Appendix}

\subsection{Additional Detail on the CAAT Graph} \label{app:CAAT_details}

\paragraph{Agent node features.}
Each agent node $A_i$ has an input feature vector $\mathbf x(A_i)\in\mathbb R^5$. It contains a Boolean indicating whether the agent is alive, remaining transit time, a Boolean indicating whether it possesses the scout capability $C_0$, and its two-dimensional position normalized to $[0,1]^2$. The remaining transit time is needed because replanning may occur while the agent is traversing an edge. In that case, the agent first finishes its committed edge, then follows its newly assigned path. The remaining transit time is normalized in the same way as the travel time edge. The agent's other capabilities are represented by its connections to capability nodes.

\paragraph{Task node features.}
Each task node $T_j$ has an input feature vector $\mathbf x(T_j)\in\mathbb R^4$. It contains the task's two-dimensional position normalized to $[0,1]^2$, a Boolean indicating whether the task has been completed, and a Boolean indicating whether its required capability is known. The required capability or its prior is represented by capability edges.

The evaluated architecture retains agent-alive and task-completion indicators inherited from an earlier problem formulation. The alive indicator is constant in the present setting, while the completion indicator distinguishes completed tasks retained in the graph. Both dead agents and completed tasks are kept in the CAAT graph.

\paragraph{Action node features.}
Each action node $K_n$ has an input feature vector $\mathbf x(K_n)\in\mathbb R^4$, whose entries correspond to execute, observe, staging, and wait, in that order. Execute, observe, and wait actions have value $1$ in their corresponding entry and $0$ elsewhere. The staging entry is $0.5$ for a single-task stage and $1$ for a two-task stage, with the other entries set to $0$. This encoding distinguishes both the action type and the number of tasks associated with a staging action.

\paragraph{Capability node initialization.}
Capability nodes represent $C_k\in\mathcal C\backslash\{C_0\}$ and share a learned 128-dimensional initial embedding. The scout capability $C_0$ is encoded as an agent feature.

\paragraph{Edge details.}
Travel-time edges connect agent--task, agent--action, and action--task pairs with finite corresponding travel times, excluding the wait action. At each planning decision, the traversal times are z-score normalized using the mean and standard deviation pooled across all valid travel time edges. This normalization gives maps with different traversal-time scales similarly scaled edge features. A travel-time edge does not imply that the agent can select the action; capability restrictions are enforced by the decoder.

\paragraph{Observation sets.}
For each non-task node of the map, we use the visibility map $\mathrm{vis}(v)$ to identify the subset of tasks with unknown requirements that can be observed from that node. Nodes with identical nonempty subsets form an observation set. Each set is represented by one observation action in the CAAT graph. Its task semantic edges connect it to the corresponding unknown tasks. For each agent, the destination is the node in the selected set with the minimum traversal time from that agent's planning position. 

\paragraph{Staging candidates.}
The staging action positions an agent near tasks whose required capabilities remain unknown. For each such task, we include up to two non-task nodes with the shortest traversal times to that task. We also construct staging locations for pairs of unknown tasks. Let $d(v,T_j)$ denote the directed shortest traversal time from node $v$ to task $T_j$. For each pair $(T_j,T_k)$, we select a non-task node minimizing $\max\{d(v,T_j),d(v,T_k)\}$. To limit the number of pair actions, we rank feasible pairs by $\max\{d(T_j,T_k),d(T_k,T_j)\}$ and retain up to $n$ pairs, where $n$ is the number of remaining unknown tasks. 

\subsection{Introduction to GNNs and Graph Transformers} \label{app:GNN_Intro}

Since our neural policy takes a graph as input, we utilize a graph neural network (GNN), also known as message passing neural network (MPNN) ~\citep{gilmer2017MPNN}. A message passing neural network is a type of neural network that creates embeddings on graphs using repeated message passing layers. 

Let $G = (\mathcal V, \mathcal E)$ be a graph, and let $\mathcal N(u)$ denote the set of neighborhood nodes for node $u \in \mathcal V$. Additionally, let $\mathbf x_u$ and $\mathbf e_{uv}$ be the features of node $u$ and edge $(u,v)$ respectively. A MPNN layer is defined as
\begin{equation} \label{eq:MPNN_layer}
    \mathbf h_u = \phi \left( \mathbf x_u, \bigoplus_{v \in \mathcal N(u)} \psi(\mathbf x_u, \mathbf x_v, \mathbf e_{uv}) \right)
\end{equation}
where $\phi, \psi$ are MLPs, and $\bigoplus$ is a permutation invariant aggregator operator such as mean or max. Intuitively, a node $u$ updates its representations by creating a message using its neighbors' embedding as $\psi(\mathbf x_u, \mathbf x_v, \mathbf e_{uv})$ and aggregating it with its own representation.

A graph attention network (GAT) is a type of MPNN that uses multi-head self-attention ~\citep{vaswani2017attention} to create the message ~\citep{velivckovic2017GAT}. A GAT layer with $K$ heads is the following:
\begin{equation}\label{eq:GAT_layer}
    \mathbf h_u = \Big\Vert_{k=1}^K \sigma \left( \sum_{v \in \mathcal N(u)} \alpha_{uv}^k W^k \mathbf x_v \right)
\end{equation}
where $\Big\Vert$ represents concatenation, $\sigma$ is an activation function such as ReLU, $\alpha_{uv}^k$ is the normalized attention coefficient for head $k$, and $W^k$ is the corresponding learnable linear transformation.

\subsection{Additional Details on the HGT} \label{app:HGT_details}

Before message passing, separate MLPs project agent, task, and action features into a common hidden dimension $d$, with parameters shared among nodes of the same type. Capability nodes start from the shared learned embedding described in Appendix~\ref{app:CAAT_details}. Let $\mathbf x_u$ denote the embedding of the current node $u$, $\mathcal N_r(u)$ denote the incoming neighbors of node $u$ under directed relation $r$, and $\mathcal R(u)$ denote its possible incoming relation types. For attention head $k$, we define
\begin{equation} \label{eq:heterogeneous_attention_score}
    s_{uv}^{r,k} = \frac{(W_Q^{r,k}\mathbf x_u)^\top(W_K^{r,k}\mathbf x_v)}{\sqrt{d_k}}, \quad v \in \mathcal N_r(u)
\end{equation}
where $d_k$ is the dimension of each attention head, and $W$ are learned matrices. For travel-time edges, an MLP $\phi_e$ is shared across all layers and all travel-time edges and produces the edge embedding
\begin{equation} \label{eq:edge_embedding}
    \mathbf z_{uv} = \phi_e(\mathbf e_{uv}).
\end{equation}
A graph transformer uses a weighted sum of the neighbors' messages as the final message. Table~\ref{tab:heterogeneous_messages} summarizes the neighbor weights and messages for the edge types. 

\begin{table*}[h]
    \centering
    \small
    \renewcommand{\arraystretch}{1.8}
    \begin{tabular}{@{}l|l|l@{}}
        \toprule
        Edge type & Neighbor weight & Message from $v \in \mathcal N_r(u)$ \\
        \midrule
        Travel Time & 
        $\displaystyle \alpha_{uv}^{r,k} = \operatorname{softmax}_{v\in\mathcal N_r(u)} \left( s_{uv}^{r,k} +(\mathbf b^{r,k})^\top\mathbf z_{uv} \right)$ &
        $\displaystyle \mathbf v_{uv}^{r,k} = W_V^{r,k}\mathbf x_v + W_E^{r,k}\mathbf z_{uv}$ \\
        Task Semantic &
        $\displaystyle \alpha_{uv}^{r,k} = \operatorname{softmax}_{v\in\mathcal N_r(u)} \left(s_{uv}^{r,k}\right)$ &
        $\displaystyle \mathbf v_{uv}^{r,k} = W_V^{r,k}\mathbf x_v$ \\
        Capability--Agent &
        $\displaystyle \alpha_{uv}^{r,k} = \operatorname{softmax}_{v\in\mathcal N_r(u)} \left(s_{uv}^{r,k}\right)$ &
        $\displaystyle \mathbf v_{uv}^{r,k} = W_V^{r,k}\mathbf x_v$ \\
        Capability--Task &
        $\displaystyle p_{uv}= p(C(T_j) = C_k)$ &
        $\displaystyle W_V^r\mathbf x_v$ \\
        \bottomrule
    \end{tabular}
    \vspace{-5pt} 
    \caption{
        Neighbor weights and message values for each edge family.
    }
    \vspace{-5pt} 
    \label{tab:heterogeneous_messages}
\end{table*}

The message $\mathbf m_u^r$ for node $u$ for relation type $r$ is defined below. 
\begin{equation} \label{eq:HGT_message_def}
    \mathbf m_u^r =
    \begin{cases}
        \sum_{v\in\mathcal N_r(u)} p_{uv}W_V^r\mathbf x_v &\text{for capability--task relations}, \vspace{0,1in}\\
        W_O^r \left( \Big\Vert_{k=1}^{K} \sum_{v\in\mathcal N_r(u)} \alpha_{uv}^{r,k}\mathbf v_{uv}^{r,k} \right) &\text{otherwise}.
    \end{cases}
\end{equation}
The node embedding is updated by the following message passing mechanism:
\begin{equation} \label{eq:heterogeneous_layer_update}
    \mathbf h_u = \mathrm{LayerNorm}_{\tau(u)} \left( \mathbf x_u + \phi_{\tau(u)} \left( \mathbf x_u, \Big\Vert_{r\in\mathcal R(u)} \mathbf m_u^r \right) \right),
\end{equation}
where $\tau(u)$ denotes the node type and $\phi_{\tau(u)}$ is a node type specific MLP. Each layer updates all node types simultaneously from the incoming embeddings. 

The messages from each relation are concatenated in a fixed order for each node type. If the neighborhood for a particular relation is empty, then its message is a zero vector. Agents, tasks, actions, and capabilities have $R_\tau=3,6,5,2$ incoming relation slots, respectively. Thus, the input dimension of $\phi_\tau$ is $(1+R_\tau)d$. The wait node has no incoming messages and is updated using its own embedding.

\paragraph{Architecture settings.}
Table~\ref{tab:HGT_architecture} specifies the network dimensions. Each MLP contains two linear layers with GELU between them. The agent, task, and action input encoders additionally apply LayerNorm after the second linear layer. The MLP decoder concatenates the final agent embedding, action embedding, and their 32-dimensional travel-time embedding. Its input dimension is therefore $2d+32=288$, and its scalar output is the agent--action logit.

\begin{table}[htbp]
    \centering
    \small
    \renewcommand{\arraystretch}{1.15}
    \begin{tabular}{@{}p{0.32\linewidth}p{0.64\linewidth}@{}}
        \toprule
        Component & Specification \\
        \midrule
        Hidden node dimension & $d=128$ \\
        Message-passing layers & $3$ \\
        Attention heads & $4$ per attention relation; $d_k=32$ \\
        Agent input encoder & $5 \rightarrow 128 \rightarrow 128$ \\
        Task input encoder & $4 \rightarrow 128 \rightarrow 128$ \\
        Action input encoder & $4 \rightarrow 128 \rightarrow 128$ \\
        Capability initialization & Shared learned vector in $\mathbb R^{128}$ \\
        Travel-time encoder & $1 \rightarrow 32 \rightarrow 32$ \\
        Node-update MLP & $128(1+R_\tau) \rightarrow 512 \rightarrow 128$ \\
        Agent--action decoder MLP & $288 \rightarrow 64 \rightarrow 1$ \\
        Dropout & $0$ \\
        \bottomrule
    \end{tabular}
    \vspace{-5pt} 
    \caption{HGT dimensions. Arrows denote linear layer dimensions.}
    \vspace{-5pt} 
    \label{tab:HGT_architecture}
\end{table}

\paragraph{Probabilities of actions with the same location.}
Different actions on the CAAT graph can point to the same location, such as staging actions associated with different tasks. The probability of assigning agent $A_i$ to that location is the sum of the probabilities of all feasible actions that lead to it. Let $\mathcal K(v)$ denote these actions for location $v$. At assignment step $s$, the resulting probability is
\begin{equation} \label{eq:destination_probability}
    q_s(A_i,v)
    = \sum_{K_n\in\mathcal K(v), (A_i, K_n) \in \mathcal F_s}
        p_s(A_i,K_n).
\end{equation}
Actions at the same location share unit assignment capacity, so selecting that location removes all of them from subsequent assignment steps.

\subsection{Additional Details on Policy pretraining} \label{app:policy_pretraining}

\paragraph{Batched policy updates.}
The batched training implementation collects $B$ episodes before each optimizer update. The EMA baseline $b$ is held fixed throughout the batch. For episode $e$ with return $R_e$, we compute $a_e=R_e-b$ and standardize these advantages across the full optimizer batch:
\begin{equation} \label{eq:batch_advantage}
    \widehat a_e = \frac{a_e-\bar a}{\sigma_a},
    \qquad
    \bar a = \frac{1}{B}\sum_{e=1}^{B}a_e,
    \qquad
    \sigma_a^2 = \frac{1}{B}\sum_{e=1}^{B}(a_e-\bar a)^2.
\end{equation}
Since $b$ is shared by all episodes, centering cancels this offset. 

Let $m_e$ be the number of replanning decisions in episode $e$. At decision $t$, $U_{e,t}$ is a random sequential assignment with conditional distribution $\pi_\theta(G_{e,t})$, and $u_{e,t}$ is the sample. We write $\pi_\theta(u_{e,t}\mid G_{e,t})=\Pr_\theta(U_{e,t}=u_{e,t}\mid G_{e,t})$ for the probability of that realization. The batch loss is
\begin{equation} \label{eq:batch_reinforce_loss}
    \mathcal L_{\mathrm{batch}}
    = -\frac{1}{B}\sum_{e=1}^{B}\frac{1}{m_e}
        \sum_{t=1}^{m_e}\left[
            \widehat a_e\log\pi_\theta(u_{e,t}\mid G_{e,t})
            +\beta H(U_{e,t}\mid G_{e,t})\right],
\end{equation}
where $H(U_{e,t}\mid G_{e,t})$ is the entropy of the conditional assignment distribution. Averaging within each episode prevents episodes with more replanning events from receiving proportionally greater weight. Note that all decisions for episode $e$ receive the same advantage $\hat a_e$ regardless of the timestep $t$.

\subsection{Baseline Planners} \label{app:baseline_planner}

The three baselines use two shared solvers: a WRP solver for discovering task requirements and an oracle solver for executing tasks with known requirements.

\paragraph{WRP solver.}
The WRP solver coordinates a set of scouts to minimize the time of the final requirement revelation. Let $\mathcal U$ be the set of tasks with unknown requirements and let $O(v)\subseteq\mathcal U$ contain those visible from node $v$ according to $\mathrm{vis}(v)$. For at most 12 unknown tasks, each scout searches augmented states $(v,S)$, where $v$ is its current map node and $S\subseteq\mathcal U$ contains the tasks observed along its route. Each map edge $(v,v')$ defines the transition
\begin{equation} \label{eq:scout_state_transition}
    (v,S)\xrightarrow{c(v,v')}(v',S\cup O(v')),
\end{equation}
where $c(v,v')$ is the traversal cost. For scout $A_i$ starting at $v_i$, let $D_i(v,S)$ denote the shortest travel time from the initial state $(v_i,O(v_i))$ to $(v,S)$. Dijkstra's algorithm determines the minimum time to observe an assigned subset $B\subseteq\mathcal U$ as
\begin{equation} \label{eq:scout_subset_cost}
    q_i(B)=\min_{\substack{v,S\\B\subseteq S}}D_i(v,S).
\end{equation}
The condition $B\subseteq S$ allows a route to observe additional tasks. A subset assignment dynamic program partitions $\mathcal U$ among the scouts to minimize the final reveal time. It uses Equation~\ref{eq:oracle_assignment} below, with $q_i$ replacing $\ell_i$ and the scouts replacing the execution agents.

Above 12 unknown tasks, each observation task is independently assigned to the scout that can observe it earliest, without accounting for other tasks assigned to that scout. Each scout's covering route is then found using weighted A*. The optimality of this assignment is no longer guaranteed. 

\paragraph{Oracle solver.}
Given a set of tasks with known requirements and a set of agents, the oracle solver finds a minimum-makespan assignment and independently executable routes. For agent $A_i$, Dijkstra's algorithm searches augmented states $(v,S)$, where $v$ is the current map node and $S$ is the set of tasks completed along the route. Traversing an edge incurs its directed travel time. Entering a compatible task adds it to $S$, whereas entering an incompatible task is prohibited. Minimizing over terminal nodes gives $\ell_i(S)$, the shortest route from the agent's starting position that completes exactly $S$. Unreachable or incompatible subsets have infinite cost, and $\ell_i(\varnothing)=0$. 

Let $\mathcal T$ be the set of tasks to complete, and let $M_i(U)$ be the minimum makespan for assigning exactly $U\subseteq\mathcal T$ to the first $i$ agents (agent order is fixed). The assignment dynamic program uses the Bellman recurrence
\begin{equation} \label{eq:oracle_assignment}
    M_i(U) = \min_{S\subseteq U}
        \max\left\{M_{i-1}(U\setminus S),\ell_i(S)\right\},
\end{equation}
with $M_0(\varnothing)=0$ and $M_0(U)=\infty$ for nonempty $U$. The resulting makespan is $M_{N_a}(\mathcal T)$. This process yields each agent's assignment, task order, and path.

When replanning during transit, both solvers use the next arrival node as the agent's starting position and add the remaining edge traversal time to the cost of any nonempty assigned subset. The agent finishes traversing its current edge before following the new path.

\paragraph{Oracle baseline.}
The oracle baseline applies the oracle solver once to all agents and tasks using the ground-truth requirements. Its makespan $M^*$ is used to normalize the learned policy's rollout reward and evaluation results.

\paragraph{Scout Then Engage.}
This policy plans in two phases. First, all scout-capable agents, including hybrids, use the WRP solver to reveal every unknown task requirement, while non-scout agents wait. Once every remaining requirement is known, the policy applies the oracle solver once to all living agents and remaining tasks using the revealed requirements, then follows the resulting execution plan.

\paragraph{Engage While Scout.}
This policy runs the WRP solver for a scouting group and the oracle solver for an execution group. The scouting group reveals unknown requirements, while the execution group completes known incomplete tasks supported by at least one of its agents. Both groups act simultaneously, and their plans are recomputed at every replanning event. The policy uses only revealed requirements and provides no staging actions. The two solvers optimize their respective subproblems independently.

Pure scouts join the scouting group, and non-scout agents join the execution group. A heuristic assigns hybrid agents according to the current known task demand. For each non-scout capability $C_k$, let $d_k$ be the number of known incomplete tasks requiring $C_k$ and $s_k$ the number of living non-scout agents possessing $C_k$. The initial capability deficit is
\begin{equation} \label{eq:hybrid_capability_deficit}
    g_k=\max\{0,d_k-s_k\}.
\end{equation}
Among the remaining hybrids, we select the agent possessing the largest number of capabilities with a positive deficit, breaking ties uniformly at random. For each capability possessed by that agent, its deficit is reduced by one. Selection repeats until all deficits are zero or no remaining hybrid can reduce a deficit. Selected hybrids join the execution group; the remaining hybrids join the scouting group while unknown requirements remain, and otherwise wait.

\subsection{Additional Details on Experiments} \label{app:detail_on_results}

\paragraph{Pretraining settings.}
Table~\ref{tab:RL_hyperparameters} lists the RL hyperparameters. Ablations use the same training budget and RL hyperparameters, changing only the components specified in Section~\ref{subsec:ablation}. For each configuration, we evaluate the checkpoint with the highest mean reward over an optimizer batch.

\begin{table}[htbp]
    \centering
    \small
    \begin{tabular}{@{}ll@{}}
        \toprule
        Parameter & Value \\
        \midrule
        Optimizer & Adam \\
        Learning rate & $3\times10^{-4}$ \\
        Episodes per optimizer update & $32$ \\
        Number of simultaneous simulations & $8$ \\
        Entropy coefficient $\beta$ & $0.01$ \\
        EMA baseline decay & $0.95$ \\
        Incompletion penalty $\lambda_I$ & $10$ \\
        Gradient norm clipping threshold & $1$ \\
        Training episodes & $12{,}800$ \\
        \bottomrule
    \end{tabular}
    \vspace{-5pt} 
    \caption{RL training hyperparameters used by the default policy and reported ablations.}
    \vspace{-5pt} 
    \label{tab:RL_hyperparameters}
\end{table}

During training, task locations are sampled uniformly without replacement from map nodes excluding the source. Task requirements are sampled independently using a uniform distribution. An agent possesses a non-scout capability $C_k$ with independent probability $0.5$, and the scout capability $C_0$ with independent probability $0.25$. If no agent possesses capability $C_k$ after this process, an agent is randomly sampled and given this capability. All agents start at node $(0,0)$ on WV.

\paragraph{Evaluation scenarios.}
The test suites combine three agent profiles with three task profiles for each $(N_a,N_t)$. Tables~\ref{tab:agent_profiles} and~\ref{tab:scenario_profiles} summarize the agent and task profiles, respectively. WV and WA test suites are identical except for the source node: $(0,0)$ for WV and $(63,63)$ for WA. At each $3 \leq N_a \leq 6$ and $5 \leq N_t \leq 9$ we test all 9 pairs of agent--task profiles, resulting in 180 total test scenarios. KTH uses the same agent and task profiles, with $N_a = 8$ and $N_t = 14$ for a total of 9 test scenarios. 

\begin{table}[htbp]
    \centering
    \small
    \begin{tabular}{@{}lp{0.70\linewidth}@{}}
        \toprule
        Agent profile & Design \\
        \midrule
        Specialist & Agents with few capabilities. \\
        Balanced hybrid & Agents with 2--3 capabilities, including hybrid scouts that can both observe and execute tasks. \\
        Scout redundant & Approximately half the agents have $C_0$, more than the other profiles which have $25\% - 33 \%$. \\
        \bottomrule
    \end{tabular}
    \vspace{-5pt} 
    \caption{Agent profiles used in the test suites.}
    \vspace{-5pt} 
    \label{tab:agent_profiles}
\end{table}

\begin{table}[htbp]
    \centering
    \small
    \begin{tabular}{@{}lp{0.70\linewidth}@{}}
        \toprule
        Task profile & Design \\
        \midrule
        Balanced & Spatially dispersed tasks with approximately balanced requirement counts. \\
        Skewed & Spatially dispersed tasks with requirements dominated by one capability on WV and WA, and two on KTH. \\
        Clustered & Tasks concentrated in multiple spatial clusters, with balanced requirements. \\
        \bottomrule
    \end{tabular}
    \vspace{-5pt} 
    \caption{Task profiles used in the test suites.}
    \vspace{-5pt} 
    \label{tab:scenario_profiles}
\end{table}

\paragraph{Result aggregation.}
Average regret for a given map is the equally weighted mean of the individual scenario's regret $M/M^*$. The error bars show the standard deviation of all the regrets for a given policy and map.

\paragraph{Planning decision time measurements.}
All runtime measurements were collected on the RTX 5090 desktop. The learned policy and its routing use the GPU; the classical solvers run on the CPU. The time measured is the time to generate a path for all agents using the current simulation state. For our method, this includes CAAT graph creation, HGT processing, decoder action assignment, and shortest path calculation. Offline map preparation and episode initialization are excluded.

\subsection{Additional Results} \label{app:additional_results}

\begin{figure}[t]
    \centering
    \includegraphics[width=\linewidth]{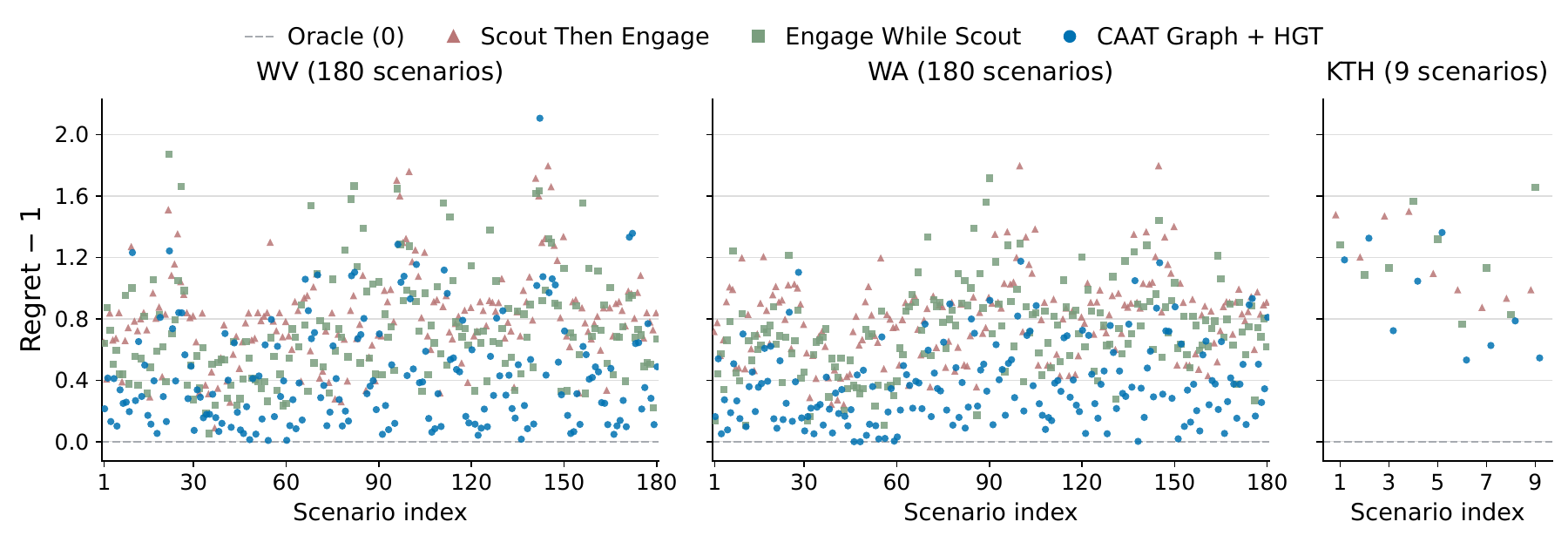}
    \vspace{-5pt} 
    \caption{Individual-scenario results on WV, WA, and KTH, plotted as $M/M^*-1$. At each scenario index, the three markers compare the policies on the same instance. The dashed line denotes the oracle.}
    \vspace{-5pt} 
    \label{fig:per_scenario_regret}
\end{figure}

Tables~\ref{tab:wv_detailed_results}--\ref{tab:kth_detailed_results} report regret by the number of agents and tasks for each map. Figure~\ref{fig:per_scenario_regret} shows the results for every individual scenario.

\begin{table}[htbp]
    \centering
    \small
    \begin{tabular}{@{}ccccc@{}}
        \toprule
        Agents & Tasks & Scout then Engage & Engage While Scout & CAAT Graph + HGT \\
        \midrule
        3 & 5 & $1.641 \pm 0.234$ & $1.654 \pm 0.198$ & $1.234 \pm 0.123$ \\
         & 6 & $1.605 \pm 0.207$ & $1.458 \pm 0.218$ & $1.306 \pm 0.203$ \\
         & 7 & $1.726 \pm 0.425$ & $1.714 \pm 0.459$ & $1.380 \pm 0.358$ \\
         & 8 & $1.839 \pm 0.344$ & $1.756 \pm 0.426$ & $1.622 \pm 0.322$ \\
         & 9 & $1.730 \pm 0.122$ & $1.443 \pm 0.187$ & $1.302 \pm 0.190$ \\
        \midrule
        4 & 5 & $1.634 \pm 0.214$ & $1.619 \pm 0.101$ & $1.172 \pm 0.123$ \\
         & 6 & $1.698 \pm 0.187$ & $1.761 \pm 0.391$ & $1.432 \pm 0.364$ \\
         & 7 & $1.821 \pm 0.129$ & $1.744 \pm 0.520$ & $1.610 \pm 0.317$ \\
         & 8 & $1.749 \pm 0.323$ & $1.761 \pm 0.317$ & $1.533 \pm 0.290$ \\
         & 9 & $1.664 \pm 0.160$ & $1.617 \pm 0.332$ & $1.325 \pm 0.217$ \\
        \midrule
        5 & 5 & $1.790 \pm 0.131$ & $1.608 \pm 0.159$ & $1.100 \pm 0.057$ \\
         & 6 & $1.904 \pm 0.338$ & $1.943 \pm 0.450$ & $1.478 \pm 0.419$ \\
         & 7 & $2.024 \pm 0.295$ & $1.949 \pm 0.296$ & $1.681 \pm 0.277$ \\
         & 8 & $1.977 \pm 0.357$ & $1.831 \pm 0.201$ & $1.651 \pm 0.271$ \\
         & 9 & $1.796 \pm 0.269$ & $1.696 \pm 0.257$ & $1.346 \pm 0.179$ \\
        \midrule
        6 & 5 & $1.812 \pm 0.129$ & $1.608 \pm 0.154$ & $1.102 \pm 0.065$ \\
         & 6 & $1.921 \pm 0.345$ & $1.898 \pm 0.409$ & $1.490 \pm 0.411$ \\
         & 7 & $2.014 \pm 0.304$ & $1.892 \pm 0.344$ & $1.850 \pm 0.538$ \\
         & 8 & $2.058 \pm 0.409$ & $1.866 \pm 0.295$ & $1.620 \pm 0.319$ \\
         & 9 & $1.862 \pm 0.265$ & $1.682 \pm 0.296$ & $1.363 \pm 0.192$ \\
        \bottomrule
    \end{tabular}
    \vspace{-5pt} 
    \caption{Regret $M/M^*$ on WV by number of agents and tasks. Entries show the mean $\pm$ standard deviation over the nine agent--task profile combinations.}
    \vspace{-5pt} 
    \label{tab:wv_detailed_results}
\end{table}

\begin{table}[htbp]
    \centering
    \small
    \begin{tabular}{@{}ccccc@{}}
        \toprule
        Agents & Tasks & Scout Then Engage & Engage While Scout & CAAT Graph + HGT \\
        \midrule
        3 & 5 & $1.716 \pm 0.246$ & $1.454 \pm 0.234$ & $1.285 \pm 0.192$ \\
         & 6 & $1.779 \pm 0.193$ & $1.671 \pm 0.139$ & $1.219 \pm 0.157$ \\
         & 7 & $1.652 \pm 0.249$ & $1.600 \pm 0.269$ & $1.282 \pm 0.158$ \\
         & 8 & $1.701 \pm 0.308$ & $1.611 \pm 0.285$ & $1.374 \pm 0.236$ \\
         & 9 & $1.616 \pm 0.233$ & $1.513 \pm 0.181$ & $1.312 \pm 0.302$ \\
        \midrule
        4 & 5 & $1.712 \pm 0.149$ & $1.591 \pm 0.202$ & $1.309 \pm 0.272$ \\
         & 6 & $1.754 \pm 0.220$ & $1.806 \pm 0.202$ & $1.292 \pm 0.152$ \\
         & 7 & $1.660 \pm 0.113$ & $1.738 \pm 0.288$ & $1.329 \pm 0.268$ \\
         & 8 & $1.758 \pm 0.277$ & $1.754 \pm 0.452$ & $1.390 \pm 0.233$ \\
         & 9 & $1.785 \pm 0.097$ & $1.864 \pm 0.472$ & $1.384 \pm 0.269$ \\
        \midrule
        5 & 5 & $1.897 \pm 0.213$ & $1.788 \pm 0.214$ & $1.314 \pm 0.225$ \\
         & 6 & $1.866 \pm 0.258$ & $1.680 \pm 0.264$ & $1.289 \pm 0.138$ \\
         & 7 & $1.816 \pm 0.228$ & $1.764 \pm 0.173$ & $1.479 \pm 0.232$ \\
         & 8 & $1.946 \pm 0.384$ & $1.768 \pm 0.302$ & $1.523 \pm 0.290$ \\
         & 9 & $1.880 \pm 0.281$ & $1.836 \pm 0.209$ & $1.501 \pm 0.253$ \\
        \midrule
        6 & 5 & $1.922 \pm 0.205$ & $1.807 \pm 0.181$ & $1.321 \pm 0.326$ \\
         & 6 & $1.867 \pm 0.258$ & $1.742 \pm 0.256$ & $1.303 \pm 0.117$ \\
         & 7 & $1.846 \pm 0.215$ & $1.721 \pm 0.133$ & $1.455 \pm 0.289$ \\
         & 8 & $1.962 \pm 0.378$ & $1.768 \pm 0.298$ & $1.566 \pm 0.314$ \\
         & 9 & $1.908 \pm 0.258$ & $1.870 \pm 0.188$ & $1.503 \pm 0.245$ \\
        \bottomrule
    \end{tabular}
    \vspace{-5pt} 
    \caption{Regret $M/M^*$ on WA by number of agents and tasks. Entries show the mean $\pm$ standard deviation over the nine agent--task profile combinations.}
    \vspace{-5pt} 
    \label{tab:wa_detailed_results}
\end{table}

\begin{table}[htbp]
    \centering
    \small
    \begin{tabular}{@{}ccccc@{}}
        \toprule
        Agents & Tasks & \shortstack{Scout Then\\Engage} & \shortstack{Engage While\\Scout} & CAAT Graph + HGT \\
        \midrule
        8 & 14 & $2.171 \pm 0.238$ & $2.196 \pm 0.282$ & $1.903 \pm 0.312$ \\
        \bottomrule
    \end{tabular}
    \vspace{-5pt} 
    \caption{Regret $M/M^*$ on KTH by number of agents and tasks. Entries show the mean $\pm$ standard deviation over the nine agent--task profile combinations.}
    \vspace{-5pt} 
    \label{tab:kth_detailed_results}
\end{table}

Tables~\ref{tab:wv_profile_results}--\ref{tab:kth_profile_results} report regret by agent and task profiles.

\begin{table}[htbp]
    \centering
    \small
    \begin{tabular}{@{}llccc@{}}
        \toprule
        Agent profile & Task profile & Scout Then Engage & Engage While Scout & CAAT Graph + HGT \\
        \midrule
        Specialist & Balanced & $2.024 \pm 0.405$ & $1.752 \pm 0.365$ & $1.558 \pm 0.493$ \\
         & Skewed & $1.885 \pm 0.276$ & $1.670 \pm 0.272$ & $1.405 \pm 0.314$ \\
         & Clustered & $2.044 \pm 0.314$ & $1.726 \pm 0.407$ & $1.469 \pm 0.394$ \\
        \midrule
        Balanced hybrid & Balanced & $1.870 \pm 0.245$ & $1.728 \pm 0.357$ & $1.483 \pm 0.350$ \\
         & Skewed & $1.618 \pm 0.240$ & $1.850 \pm 0.358$ & $1.337 \pm 0.229$ \\
         & Clustered & $1.825 \pm 0.164$ & $1.813 \pm 0.333$ & $1.452 \pm 0.314$ \\
        \midrule
        Scout redundant & Balanced & $1.733 \pm 0.257$ & $1.720 \pm 0.366$ & $1.440 \pm 0.347$ \\
         & Skewed & $1.589 \pm 0.198$ & $1.537 \pm 0.278$ & $1.272 \pm 0.208$ \\
         & Clustered & $1.730 \pm 0.221$ & $1.729 \pm 0.309$ & $1.454 \pm 0.339$ \\
        \bottomrule
    \end{tabular}
    \vspace{-5pt} 
    \caption{Regret $M/M^*$ on WV by agent and task profiles. Entries show the mean $\pm$ standard deviation over the 20 agent--task count combinations.}
    \vspace{-5pt} 
    \label{tab:wv_profile_results}
\end{table}

\begin{table}[htbp]
    \centering
    \small
    \begin{tabular}{@{}llccc@{}}
        \toprule
        Agent profile & Task profile & Scout Then Engage & Engage While Scout & CAAT Graph + HGT \\
        \midrule
        Specialist & Balanced & $1.994 \pm 0.321$ & $1.730 \pm 0.327$ & $1.507 \pm 0.295$ \\
         & Skewed & $1.867 \pm 0.264$ & $1.586 \pm 0.306$ & $1.276 \pm 0.241$ \\
         & Clustered & $1.993 \pm 0.303$ & $1.753 \pm 0.286$ & $1.386 \pm 0.318$ \\
        \midrule
        Balanced hybrid & Balanced & $1.677 \pm 0.198$ & $1.809 \pm 0.202$ & $1.434 \pm 0.251$ \\
         & Skewed & $1.689 \pm 0.242$ & $1.647 \pm 0.203$ & $1.251 \pm 0.154$ \\
         & Clustered & $1.873 \pm 0.079$ & $1.799 \pm 0.173$ & $1.412 \pm 0.211$ \\
        \midrule
        Scout redundant & Balanced & $1.674 \pm 0.194$ & $1.698 \pm 0.276$ & $1.353 \pm 0.221$ \\
         & Skewed & $1.660 \pm 0.223$ & $1.633 \pm 0.368$ & $1.262 \pm 0.194$ \\
         & Clustered & $1.793 \pm 0.210$ & $1.802 \pm 0.262$ & $1.462 \pm 0.274$ \\
        \bottomrule
    \end{tabular}
    \vspace{-5pt} 
    \caption{Regret $M/M^*$ on WA by agent and task profiles. Entries show the mean $\pm$ standard deviation over the 20 agent--task count combinations.}
    \vspace{-5pt} 
    \label{tab:wa_profile_results}
\end{table}

\begin{table}[htbp]
    \centering
    \small
    \begin{tabular}{@{}llccc@{}}
        \toprule
        Agent profile & Task profile & Scout Then Engage & Engage While Scout & CAAT Graph + HGT \\
        \midrule
        Specialist & Balanced & $2.479$ & $2.284$ & $2.184$ \\
         & Skewed & $2.203$ & $2.086$ & $2.325$ \\
         & Clustered & $2.470$ & $2.131$ & $1.723$ \\
        \midrule
        Balanced hybrid & Balanced & $2.500$ & $2.564$ & $2.045$ \\
         & Skewed & $2.096$ & $2.319$ & $2.362$ \\
         & Clustered & $1.990$ & $1.764$ & $1.532$ \\
        \midrule
        Scout redundant & Balanced & $1.875$ & $2.131$ & $1.627$ \\
         & Skewed & $1.934$ & $1.827$ & $1.787$ \\
         & Clustered & $1.990$ & $2.656$ & $1.545$ \\
        \bottomrule
    \end{tabular}
    \vspace{-5pt} 
    \caption{Regret $M/M^*$ on KTH by agent and task profiles. Each entry is one scenario with eight agents and fourteen tasks.}
    \vspace{-5pt} 
    \label{tab:kth_profile_results}
\end{table}

\end{document}